\documentclass[prl,reprint,superscriptaddress,longbibliography]{revtex4-1}
\usepackage[colorlinks=true,allcolors=blue,breaklinks=true]{hyperref}
\usepackage{graphicx}
\usepackage{siunitx}
\usepackage[T1]{fontenc}
\usepackage{lmodern}

\begin{document}
\title{Thermal transport in nanoelectronic devices cooled by on-chip magnetic refrigeration}

\author{S.~Autti}
\email{s.autti@lancaster.ac.uk}
\affiliation{Department of Physics, Lancaster University, Lancaster, LA1 4YB, UK.}

\author{F.~C.~Bettsworth}
\affiliation{Department of Physics, Lancaster University, Lancaster, LA1 4YB, UK.}

\author{K.~Grigoras}
\affiliation{VTT Technical Research Centre of Finland Ltd, P.O. Box 1000, 02044 VTT, Espoo, Finland.}

\author{D.~Gunnarsson}
\altaffiliation{Present address: BlueFors Cryogenics Oy, Arinatie 10, 00370 Helsinki, Finland.}
\affiliation{VTT Technical Research Centre of Finland Ltd, P.O. Box 1000, 02044 VTT, Espoo, Finland.}

\author{R.~P.~Haley}
\affiliation{Department of Physics, Lancaster University, Lancaster, LA1 4YB, UK.}

\author{A.~T.~Jones}
\altaffiliation{Present address: ISIS Neutron \& Muon Source, Rutherford Appleton Laboratory, Chilton, Didcot OX11 0QX, Oxfordshire, UK.}
\affiliation{Department of Physics, Lancaster University, Lancaster, LA1 4YB, UK.}

\author{Yu.~A.~Pashkin}
\affiliation{Department of Physics, Lancaster University, Lancaster, LA1 4YB, UK.}

\author{J.~R.~Prance}
\affiliation{Department of Physics, Lancaster University, Lancaster, LA1 4YB, UK.}

\author{M.~Prunnila}
\affiliation{VTT Technical Research Centre of Finland Ltd, P.O. Box 1000, 02044 VTT, Espoo, Finland.}

\author{M.~D.~Thompson}
\affiliation{Department of Physics, Lancaster University, Lancaster, LA1 4YB, UK.}

\author{D.~E.~Zmeev}
\affiliation{Department of Physics, Lancaster University, Lancaster, LA1 4YB, UK.}




\begin{abstract}
On-chip demagnetization refrigeration has recently emerged as a powerful tool for reaching microkelvin electron temperatures in nanoscale structures. The relative importance of cooling on-chip and off-chip components and the thermal subsystem dynamics are yet to be analyzed. We study a Coulomb blockade thermometer with on-chip copper refrigerant both experimentally and numerically, showing that dynamics in this device are captured by a first-principles model. Our work shows how to simulate thermal dynamics in devices down to microkelvin temperatures, and outlines a recipe for a low-investment platform for quantum technologies and fundamental nanoscience in this novel temperature range. 
\end{abstract}

\maketitle

Easy access to millikelvin temperatures has driven the expansion of quantum research and technologies for the past decade. Understanding microkelvin physics has the potential to become the next field-defining development. Until recently, the lowest electron temperature reached by external refrigeration in micro- and nanoelectronic devices was $\approx \SI{4}{\milli\kelvin}$~\cite{Xia2000,Samkharadze2011,Bradley2016,jones2020progress}. Electrons in bulk metals are routinely cooled to much lower temperatures by nuclear demagnetization refrigeration, but the thermal coupling to a device via an insulating structure vanishes rapidly as temperature approaches the microkelvin regime. This limitation to reaching microkelvin on-chip electron temperatures has recently been overcome using two different approaches: cooling of metallic contacts by immersion in $^\mathrm{3}$He~\cite{Levitin2022} and incorporation of miniaturized on-chip demagnetization elements ~\cite{Bradley2017,Palma2017a,Yurttaguel2019}. In combination with demagnetization cooldown of all external structures in direct contact with the target device, the latter approach has produced on-chip demagnetization with hours of hold time at microkelvin temperatures~\cite{Sarsby2019,samani2022microkelvin}. However, the simultaneous demagnetization of multiple coolants makes it challenging to gain insight into subsystem temperatures that cannot be directly measured, hiding the dynamic details from view. Traditional bulk demagnetization infrastructure can also be prohibitively voluminous and expensive for many applications.

\begin{figure}[hbt!]
\includegraphics[width=1\linewidth]{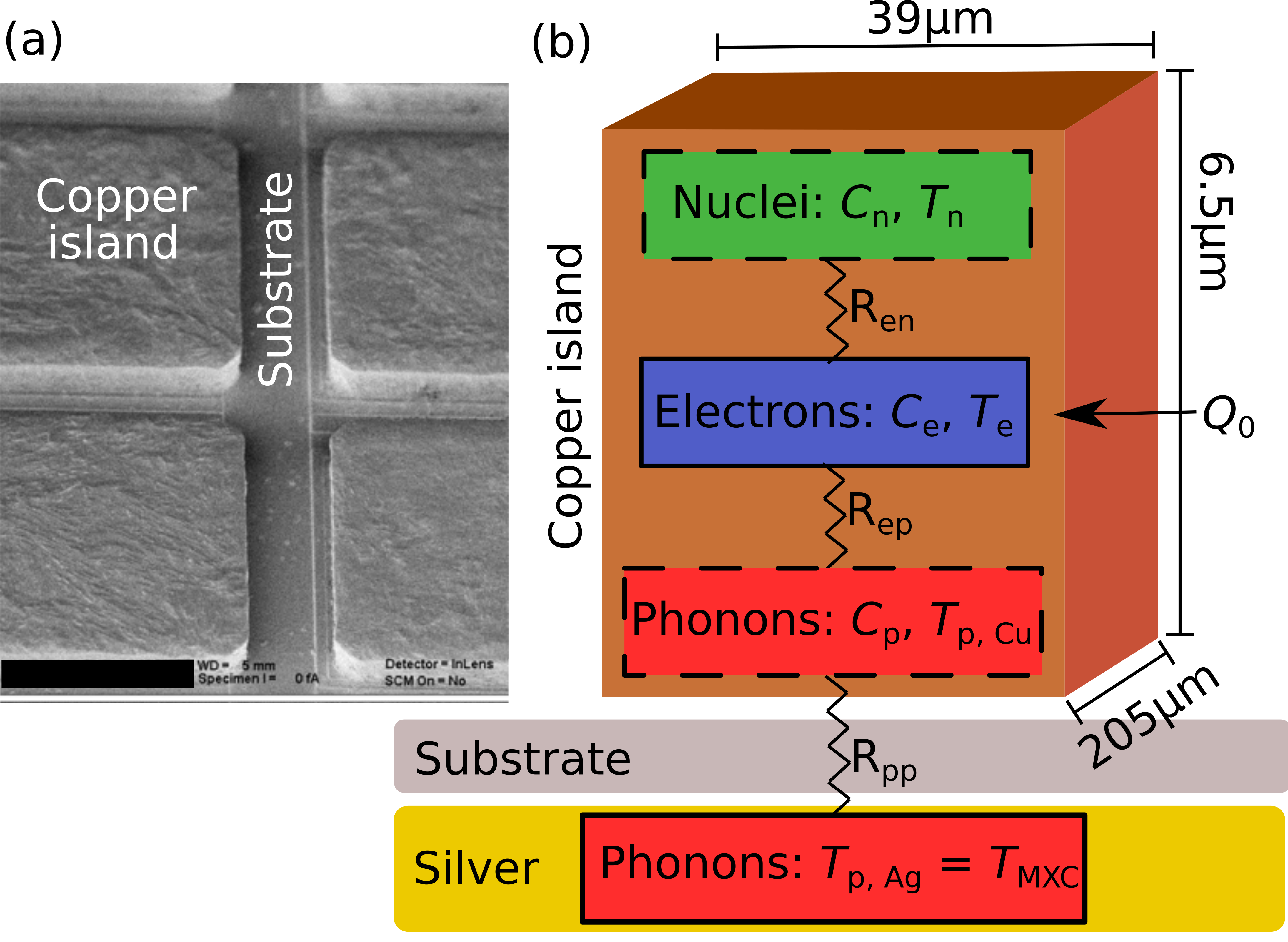}%
\caption{(a) A micrograph showing parts of four CBT demagnetization islands. The copper islands are on a silicon substrate, seen in the mircrograph between the islands. The CBTs consist of an array of $32\times20$ copper islands, each with dimensions as shown in panel (b), connected via tunnel junctions \cite{Bradley2017,samani2022microkelvin}. The black scale bar corresponds to \SI{25}{\micro\meter}. (b) Schematic illustration of the on-chip demagnetization system as implemented in the simulations. The simulation concerns one isolated copper island. The copper island is placed on a layered substrate, thermalized via an underlying silver platform to the dilution cryostat's mixing chamber at temperature $T_\mathrm{MXC}$. Phonons at temperature $T_\mathrm{p}$ in the copper island are coupled to phonons at temperature $T_\mathrm{MXC}$ in the silver layer via the combined Kapitza resistance $R_\mathrm{pp}$ of the substrate layers. The copper phonons couple to copper electrons at temperature $T_\mathrm{e}$ via electron-phonon coupling resistance $R_\mathrm{ep}$. Copper nuclei are coupled to the electrons via $R_\mathrm{en}$. Each copper subsystem carries an associated heat capacity ($C_\mathrm{e}, C_\mathrm{n}, C_\mathrm{p}$). The external heat leak to the electron system is labeled $Q_0$. Note that the copper island dimensions are not to scale for illustrational reasons.  \label{fig:schematic}}
\end{figure}

\begin{figure}
\includegraphics[width=1\linewidth]{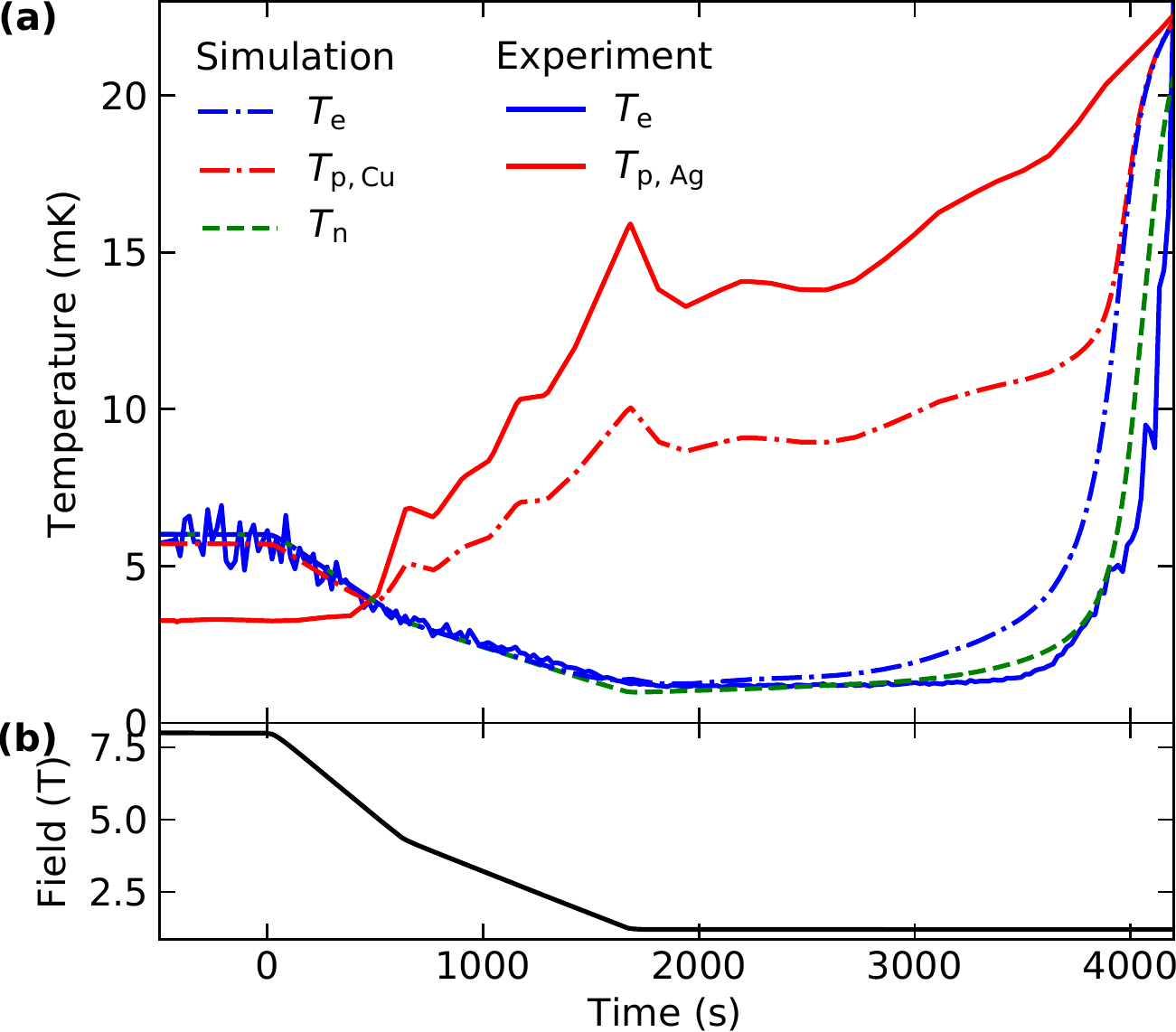}%
\caption{Gate CBT demagnetization experiment and simulation. (a) The measured gCBT electron temperature (blue solid line) and the measured mixing chamber temperature ($T_\mathrm{p,Ag}$, red solid line) level out during the pre-cooling before the demagnetization begins as shown in Fig.~\ref{fig:demag2_precool} in \cite{SM_note}. The remaining temperature difference corresponds to a fitted constant heat leak of $Q_0\approx\,$\SI{0.1}{\femto\watt}. The simulation begins at $t=-300$\,ks with initial temperatures set equal to the measured $T_\mathrm{e}$. The simulated copper nuclear temperature (green dashed line), electron temperature (blue dashed line), and phonon temperatures (red dashed line) converge around $6$\,mK during the pre-cooling. $T_\mathrm{p,Cu}$ remains $\approx0.3$\,mK lower than $T_\mathrm{e}$ and $T_\mathrm{n}$ due to the direct heat leak $Q_0$ to the electron system. The substrate phonon temperature is assumed to follow the mixing chamber temperature. During the demagnetization, the decreasing copper nuclear temperature cools the electrons. The measured electron temperature reaches $T_\mathrm{e} \approx1.1$\,mK, well in line with the simulated electron temperature. The mixing chamber temperature increases during and after the demagnetization due to eddy current heating in the metallic structures of the cryostat. The accumulated temperature-dependent heat leak from the substrate phonons in the simulation yields a hold time within 200~s ($10\%$) of that observed in the experiment. The total Kapitza resistance is $R_\mathrm{pp}=0.8 \times 10^{-2}$\SI{}{\kelvin^4 \meter^2/\watt}. (b) The magnetic field is set to \SI{8}{\tesla} during the refrigerator pre-cooling at $t<0$, and is reduced to \SI{1.2}{\tesla} in 1600~s during the demagnetization, and then held constant.\label{fig:demag2}}
\end{figure}

In this Article, we investigate a minimalist copper on-chip demagnetization system in a helium-bath dilution refrigerator. We report the coldest on-chip-only demagnetization cooldown experiment to date, reaching a device electron temperature  $T_\mathrm{e}\approx1\,$mK. The demagnetization dynamics are well captured by a first-principles thermal model, enabling analysis and predictions beyond experimental limitations. We show that the demagnetization base temperature and hold time are determined by the heat leak carried by the substrate phonons, originating from the grounded metal layer under the substrate. We predict that holding the metal layer at a manageable \SI{3}{\milli\kelvin} would allow reaching hours of hold time at microkelvin temperatures. Perhaps surprisingly, this can also be achieved with ``hot'' electrical connections to the demagnetization island or, alternatively, a contact resistance as low as \SI{50}{\ohm}. We also predict that engineering the substrate to optimize the phonon heat leak allows achieving microkelvin on-chip temperatures even in a commercial \SI{10}{\milli\kelvin} cryostat.

We study the on-chip demagnetization process using devices consisting of copper islands as illustrated in Fig.~\ref{fig:schematic}a. The islands are attached to a layered substrate (see \cite{SM_note}), placed on an underlying silver platform. The silver platform is well thermalized to the mixing chamber of the dilution refrigerator at temperature $T_\mathrm{MXC}$ via silver sinters \cite{franco1984properties,PhysRevB.102.064508}. Thus, the phonon temperature in the silver layer $T_\mathrm{p,Ag}=T_\mathrm{MXC}$. $T_\mathrm{MXC}$ is measured directly using a vibrating wire thermometer immersed in the liquid $^\mathrm{3}$He/$^\mathrm{4}$He mixture \cite{Bradley1990,Pentti2011}. The copper contains three independent thermal reservoirs, phonons at temperature $T_\mathrm{p,Cu}$, electrons at temperature $T_\mathrm{e}$, and nuclear spins at $T_\mathrm{n}$. The phonon and nuclear temperatures cannot be directly measured. The electron temperature in an array of such copper islands, connected by tunnel junctions, can be measured in-situ by operating the device as a Coulomb Blockade Thermometer (CBT)~\cite{Pekola1994}. Each CBT tunnel junction resistance is of the order of \SI{\sim 10}{\kilo\ohm}, as required for the operation of the CBT. This means that the tunnel junctions do not provide a significant thermal link between adjacent islands and the array can thermally be described as a collection of isolated copper islands, as shown in Fig.~\ref{fig:schematic}b.

The experimental results described below were obtained from two different CBTs with on-chip copper refrigerant. Both were made following the same design and fabrication process as in our previous studies~\cite{Bradley2016,Bradley2017,Prunnila2010}. The first device is a ``junction CBT'' (jCBT), meaning that the island capacitance is dominated by the capacitance of the tunnel junctions between islands. For the second device, the capacitance of the metal islands was increased by a factor of roughly $6$ by coating the islands with a dielectric layer followed by a layer of metal \cite{samani2022microkelvin}. This reduces the island charging energy $E_\mathrm{C}$ and enables operation down to at least $\approx \SI{300}{\micro\kelvin}$~\cite{samani2022microkelvin}. CBT devices where the capacitance of each island is dominated by its capacitance to a gate electrode or ground rather than by its coupling to neighboring islands are referred to as “gate CBTs” (gCBT) \cite{samani2022microkelvin}. After calibration as a function of bias voltage, the CBT electron temperature can be inferred from the zero-bias differential conductance of the device as detailed in \cite{SM_note}.

\begin{figure}
\includegraphics[width=1\linewidth]{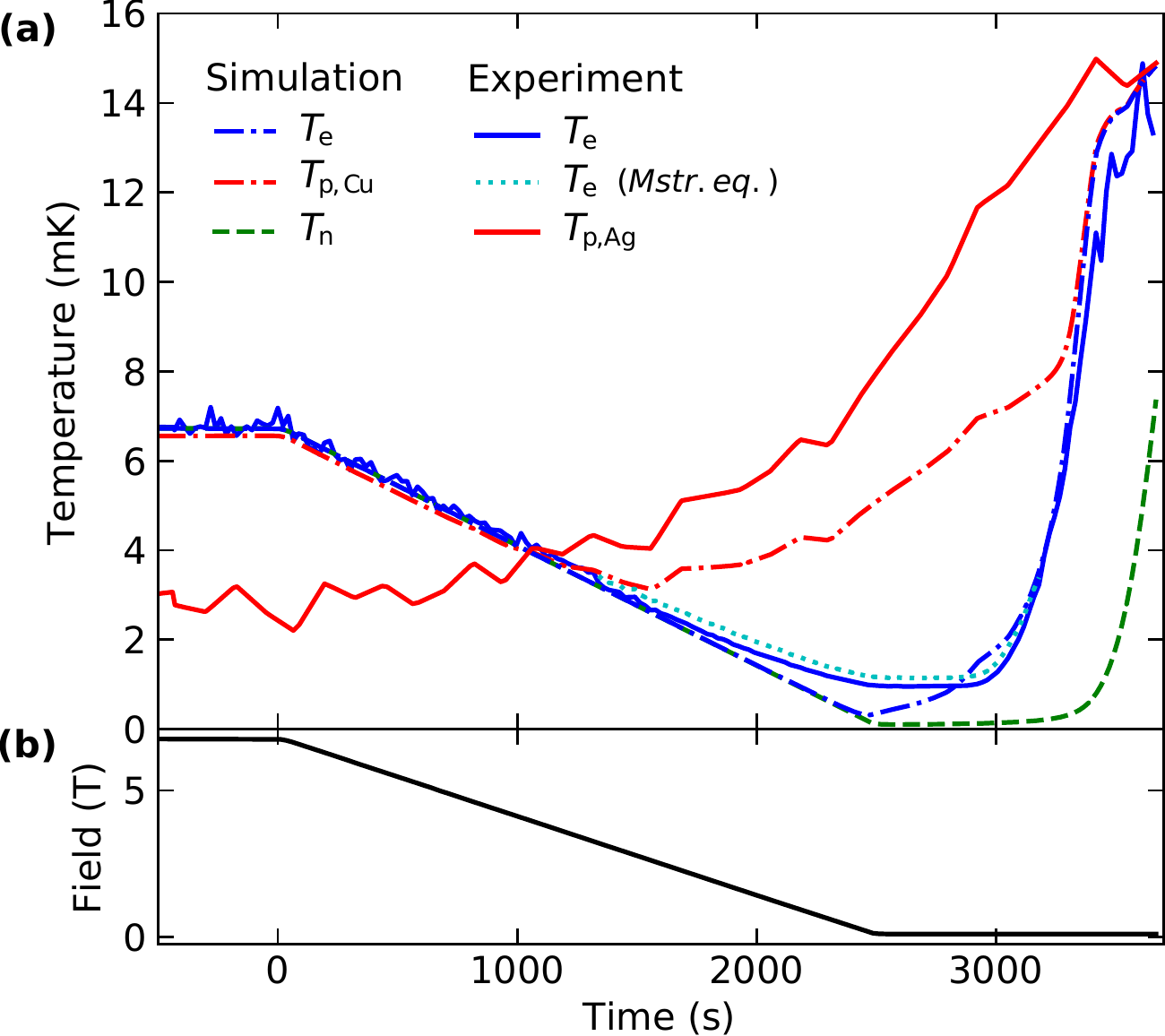}%
\caption{Junction CBT demagnetization experiment and simulation. (a) The measured jCBT electron temperature is shown by the solid blue line and the measured mixing chamber temperature ($T_\mathrm{p,Ag}$) by the solid red line. The cyan dotted line shows $T_\mathrm{e}$ given by a CBT calibration based on a master equation model where this significantly deviates from a more accurate Markov chain Monte Carlo (MCMC) model. The MCMC model accounts for the effect of random CBT island offset charges at low temperatures. To produce the solid blue line, we have converted measured conductances to CBT electron temperatures using Fig.~8 in \cite{Yurttaguel2021} as a look-up table \cite{Yurttaguel2021zenodo}. The measured electron temperature saturates as $T_\mathrm{e}=\SI{1.0\pm 0.1}{\milli\kelvin}$, caused by the AC excitation used to read out the temperature (see Fig.~\ref{fig:supp_jCBTsaturation} in \cite{SM_note}). Thus, the base temperature in the experiment is likely well below 1\,mK but it cannot be directly measured. The simulation begins at $-200$\,ks with initial temperatures set equal to the measured $T_\mathrm{e}$. The demagnetization starts at $t=0$. The decreasing nuclear temperature cools the electrons in good agreement with the experiment. The simulated electron temperature decreases to \SI{340}{\micro\kelvin}. The mixing chamber temperature increases during the demagnetization due to eddy current heating in the metallic structures of the cryostat. This increase causes the copper island to heat sharply after the demagnetization ends, before the nuclear heat capacity is exhausted, because the electron-nuclei coupling is no longer sufficient for draining the heat leak from the phonons. The fitted constant heat leak is $Q_0\approx$\,\SI{0.1}{\femto\watt}, and the Kapitza resistance is $R_\mathrm{pp}=1.5 \times 10^{-2}$\SI{}{\kelvin^4 \meter^2/\watt}. (b) The magnetic field is set to \SI{6.75}{\tesla} during the pre-cooling at $t<0$, and is then reduced linearly to \SI{100}{\milli\tesla} in 2500\,s. \label{fig:demag}}
\end{figure}

\begin{figure}
\includegraphics[width=1\linewidth]{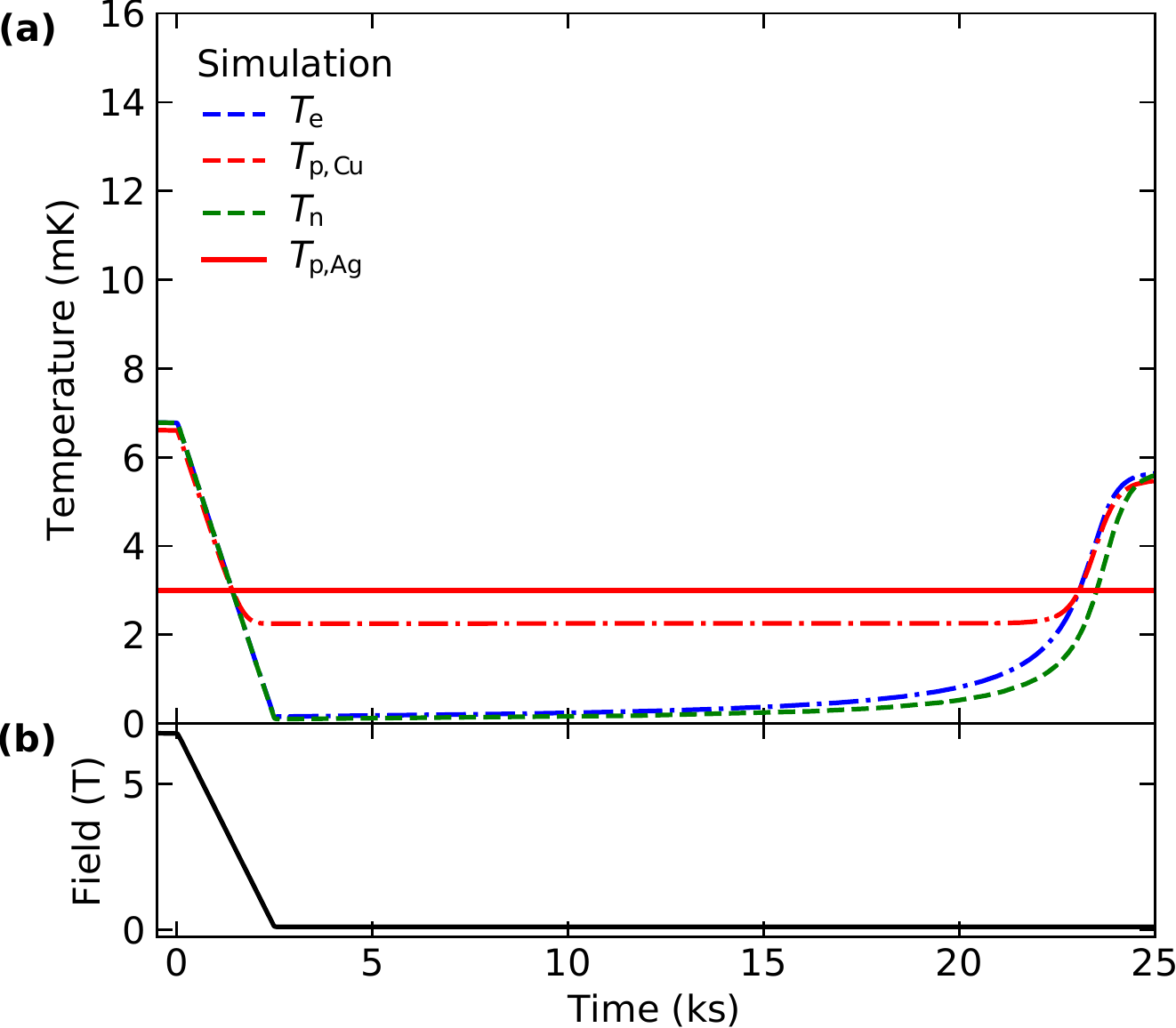}%
\caption{Predicted CBT demagnetization performance with \SI{3}{\milli\kelvin} substrate temperature. (a) If the mixing chamber temperature, equal to to the silver platform temperature (solid red line), is fixed at \SI{3}{\milli\kelvin}, the demagnetization electron temperature (blue dash line) decreases to $T_\mathrm{e}=$\,\SI{160}{\micro\kelvin}, staying under \SI{1}{\milli\kelvin} for over 5 hours. The copper phonon temperature (red dash line) levels out at \SI{2.5}{\milli\kelvin} during the simulation, and the heat leak from the phonons to the electrons falls to \SI{10}{\atto\watt}. Therefore, the constant heat leak $Q_0=0.1$\,fW determines the hold time. The simulated copper nuclear temperature is shown by the green dash line. (b) The magnetic field follows the same profile as in the experiment shown in Fig.~\ref{fig:demag}. \label{fig:demag_speculation}}
\end{figure}

Using nuclear demagnetization cooling, the CBT islands can be cooled to temperatures below those reachable by continuous cooldown. Demagnetization experiments start by ramping the magnetic field $B$ to its maximum value (here $8$~T), and then waiting for the dilution refrigerator and the CBT's electrons to thermalize. After a few days, the CBT electrons reach their equilibrium temperature $T_\mathrm{e}\approx \SI{7}{\milli\kelvin}$, and the mixing chamber temperature stabilizes at $T_\mathrm{MXC}\approx 3$~mK. The data measured using the gate CBT, shown in Figure \ref{fig:demag2}, starts in this pre-cooled configuration. The pre-cooling is shown in Fig.~\ref{fig:demag2_precool} in \cite{SM_note}.

The entropy of the nuclear system is purely a function of $B/T_\mathrm{n}$. For an ideal adiabatic demagnetization process this ratio remains constant~\cite{Pobell2007}, causing $T_\mathrm{n}$ to decrease proportional to the change in the field magnitude. Copper electrons are then cooled via the coupling between the electrons and the nuclei. After the pre-cooling, the field is swept down to $1.2$~T while recording the zero bias conductance of the CBT and continuously adjusting the DC bias to ensure zero bias is maintained. The minimum measured electron temperature reached in Fig.~\ref{fig:demag2} is $T_\mathrm{e}\approx 1.1$\,mK. Systematic uncertainties in the measured temperatures are discussed in \cite{SM_note}, while measurement noise is insignificant at $T_\mathrm{e}\leq 4$\,mK due to narrowing of the CBT conductance dip.

We simulate the subsystem temperature evolution in a single copper island. The copper nuclear temperature is assumed to change adiabatically as a function of $B$. The nuclear, electron, and phonon temperatures then evolve as determined by their associated temperature-dependent heat capacities and couplings, and according to associated heat leaks to the electrons and phonons (see Fig.~\ref{fig:schematic}b). The parameter values and functional dependencies are taken from literature \cite{jones2020progress}. Additionally, the estimated total Kapitza resistance through the layered substrate, $R_\mathrm{pp}$, agrees with the values extracted from the pre-cooling data for each CBT device. All these aspects are detailed and substantiated in \cite{SM_note}. 

The temperature difference between the mixing chamber and the copper electrons at the end of the pre-cooling corresponds to a constant heat leak of $Q_0\approx\,$\SI{0.1}{\femto\watt} to the electron system. This is used as an input parameter for the demagnetization simulation. This heat leak is likely caused by small vibrations of the CBT holder in the magnetic field, resulting in eddy current heating~\cite{Todoshchenko2014} when the full magnetic field is applied. This means that $Q_0$ probably decreases when the field is decreased, but we ignore that for simplicity. Even in the absence of any heat leak the electron temperature would only very slowly decrease below \SI{4}{\milli\kelvin} during the pre-cooling because the nuclear heat capacity is very large in $B=8$\,T. 

The simulation shown in Fig.~\ref{fig:demag2} starts at $t=-300$\,ks, first replicating three days of pre-cooling data (Fig.~\ref{fig:demag2_precool}). The demagnetization is simulated from the initial state obtained this way. During the demagnetization, the weakening thermal link between phonons and electrons allows the electrons to be cooled below the phonon temperature. Eddy currents generated in the refrigerator support structures heat up the mixing chamber with a delay, leading to increasing $T_\mathrm{p, Ag}$. The copper phonons are kept at an elevated temperature by the heat leak via $R_\mathrm{pp}$. The simulated $T_\mathrm{e}$ during the demagnetization is in nearly perfect agreement with the experiment. After about 2000~s at the final magnetic field, the electron temperature starts to increase rapidly. Remarkably, the simulation replicates this hold time within $\sim200\,$s and attributes it to the integrated heat leak from the substrate phonons. This justifies the assumption that the heat transport through the layered substrate is described by the combined Kapitza resistance $R_\mathrm{pp}$. Note that if we simulate the CBT demagnetization adding $3\,$mK to $T_\mathrm{p, Ag}$ so that $T_\mathrm{p, Ag}\approx6\,$mK at the end of the pre-cooling and $T_\mathrm{p, Ag}\approx25\,$mK at the end of the demagnetization, the hold time is halved. On the other hand, changing $R_\mathrm{pp}$ by 50\% does not significantly change the simulation outcome (see \cite{SM_note}). That is, the exact value of $R_\mathrm{pp}$ unimportant and, instead, the temperature difference across the substrate defines the resulting heat transport. Note also that $Q_0$ plays no role in determining the hold time. Therefore, the thermal subsystem temperatures evolve as predicted by first-principles expressions in this novel temperature range. 

The minimum $T_\mathrm{e}$ is determined by the balance of the phonon heat leak against the coupling between the nuclei and electrons. The electron-nuclei heat flow does not increase significantly if the nuclei are made yet colder. If we continue the demagnetization below 1.2~T, there is practically no improvement in the minimum $T_\mathrm{e}$ reached in either the simulation or the experiment owing to the substrate heat leak. 

We can mitigate the dynamic increase in $T_\mathrm{MXC}$ by starting from a lower initial $B=6.75$\,T. Here we used a "junction" CBT, which loses sensitivity to electron temperature below $T_\mathrm{e}\approx1$\,mK (see \cite{SM_note}), but has a larger Kapitza resistance and is thus better protected from the substrate heat leak. This allows us to ramp the field down to $100\,$mT to extract all cooling capacity from the nuclei. The experimental outcome is compared with a simulation in Fig.~\ref{fig:demag}. 

The lowest measured CBT electron temperature is $T_\mathrm{e}\approx1.0\pm0.1$~mK. This reading is caused by the AC excitation used to read out the zero-bias conductance of the device as explained in \cite{SM_note}. In the simulation, the minimum electron temperature is $\approx$\,\SI{340}{\micro\kelvin}. It thus seems likely that also the experiment cools well into the microkelvin temperature regime, but obtaining conclusive evidence requires using a device where such readout limitations are absent. We note that while the gCBT design removes this issue, the resulting reduction in the Kapitza resistance makes microkelvin temperatures harder to reach with elevated phonon temperatures. The hold time in the junction CBT measurement is reduced as compared with Fig.~\ref{fig:demag2} because the nuclear heat capacity is proportional to $B^2$. That is, the simulation reproduces the experimental data nearly perfectly where the experimental data is reliable. Combining this with the analysis of the gate CBT, we conclude that the simulation model can be used to predict the dynamic demagnetization performance in conditions yet beyond experimental reach, and not only where the predicted hold time is long or the substrate temperature stable.

The only relevant heat leak contribution is that carried by the substrate phonons. If we fix the silver layer phonon temperature in the simulation to 3\,mK as shown in Fig.~\ref{fig:demag_speculation}, a demagnetization otherwise identical to that in Fig.~\ref{fig:demag} results in a minimum electron temperature of \SI{160}{\micro\kelvin} and a hold time of several hours under 1\,mK. The hold time in this configuration is limited by $Q_0$, which is likely to decrease when $B$ is reduced owing to decreased eddy current heating in the CBT. Note that electronic heat conductivity along the electric connection to the demagnetization islands (resistance $\gg$\SI{1}{\kilo\ohm}) is negligible assuming the wiring is at the same temperature as the silver platform. This implies that external demagnetization to cool the electrons in the wiring \cite{Sarsby2019,samani2022microkelvin} should not be necessary for reaching microkelvin device temperatures. 

The 3\,mK silver layer temperature could be achieved by optimizing the performance of the dilution refrigerator, by replacing the silver layer with copper that will demagnetize alongside the CBT, or by using a sub-mK refrigerator platform \cite{nyeki2022high,schmoranzer2020design}. However, if we increase the Kapitza resistance by two orders of magnitude as discussed in the Supplemental Material (Fig.~\ref{fig:demag_speculation_highK}), sub-mK electron temperatures can be achieved even if the substrate can only be cooled to 10\,mK. This would allow low-investment access to sub-mK temperatures even in a commercial cryogen-free dilution refrigerator. 

The on-chip cooldown performance and thermal dynamics of diverse new devices can be predicted using the simulation techniques presented in this Letter. For example, the electric contact resistance to the demagnetized metal could be lowered to $\sim$\,\SI{50}{\ohm} without compromising the demagnetization performance, provided the electron temperature of the wiring is kept at 3\,mK. Low contact resistances are essential in studying for example two-dimensional electron gases in semiconductors, nanomechanical resonators, and quantum circuits, all of which have previously been subjected to microkelvin bulk cooling techniques \cite{cattiaux2021macroscopic,PhysRevA.101.012336,Levitin2022}. On-chip refrigeration could allow access to temperatures deep in the microkelvin regime in devices like these using only a standard dilution refrigerator platform. We emphasize that the technique can be used to cool also phonons in the materials in contact with the demagnetization metal. Thus, on-chip magnetic refrigeration can be applied to a broad range of devices for the purposes of novel fundamental nanoscience and quantum technologies.

\begin{acknowledgments}
All the data in this Letter are available at https://doi.org/10.17635/lancaster/researchdata/xxx, including descriptions of the data sets. The simulation codes can be obtained from the corresponding author upon reasonable request.

This research is supported by the U.K.\@ EPSRC (EP/K01675X/1, EP/N019199/1, EP/P024203/1, EP/L000016/1, and EP/W015730/1), the European FP7 Programme MICROKELVIN (228464), the European Union's Horizon 2020 research and innovation programme (European Microkelvin Platform 824109, and EFINED 766853), by the Academy of Finland through the Centre of Excellence program (projects 336817 and 312294), and by Business Finland through QuTI-project (40562/31/2020). 
S.A.\@ acknowledges financial support from the Jenny and Antti Wihuri Foundation via the Council of Finnish Foundations. M.D.T acknowledges financial support from the Royal Academy of Engineering (RF\textbackslash 201819\textbackslash 18\textbackslash 2).
\end{acknowledgments}

\clearpage

\renewcommand{\thefigure}{S\arabic{figure}} 
\setcounter{figure}{0}
\renewcommand{\theequation}{S\arabic{equation}} 
\setcounter{equation}{0}

\section*{Supplemental Material}

\subsection*{CBT as a primary thermometer}

We measure the differential conductance $G$ of the CBTs as a function of DC bias using a current driven, four terminal lock-in measurement. In the weak Coulomb blockade limit ($k_\mathrm{B}T \sim E_\mathrm{C}$, where $E_\mathrm{C}$ is the charging energy of the islands), $G$ develops a temperature dependent conductance dip around zero bias that can be used for primary thermometry of the electron temperature in the CBT islands~\cite{Pekola1994,Farhangfar1997}. The electron temperature is given by the full width at half minimum of the conductance dip, $V_{1/2} \approx 5.439Nk_\mathrm{B}T_\mathrm{e}/e$, where $N$ is the number of junctions in series in the array~\cite{Pekola1994}. Alternatively, a master equation (ME) model of electron tunneling in the CBT can be fitted to curves measured at several temperatures~\cite{Farhangfar1997,Bradley2016} to determine the characteristic tunnel junction resistance $R_\mathrm{T}$ and island capacitance. From these two values, the model can then be used to convert the zero-DC-bias conductance $G_0$ to electron temperature. This self-calibration allows the electron temperature to be determined in ``secondary mode'' where only the zero-bias conductance is measured. This is faster than measuring the full conductance curve and minimizes overheating that occurs when a DC bias is applied~\cite{Farhangfar1997,Bradley2016}. The ME model is valid in the ``universal regime'', which extends down to a conductance suppression of $G_0/G_\mathrm{T} \approx 0.65$, where $G_\mathrm{T}$ is the asymptotic conductance at high bias~\cite{samani2022microkelvin}. Below this temperature, the behavior of jCBTs and gCBTs diverge and both start to disagree with the ME model. Instead, a Markov chain Monte Carlo model can be used to predict the conductance~\cite{Yurttaguel2021,samani2022microkelvin}. The experiments described in this Letter use both CBTs in secondary mode and almost entirely in the universal regime. 

\subsection*{CBT calibration, uncertainty and saturation}

\begin{figure}[htb!]
    \centering
    \includegraphics{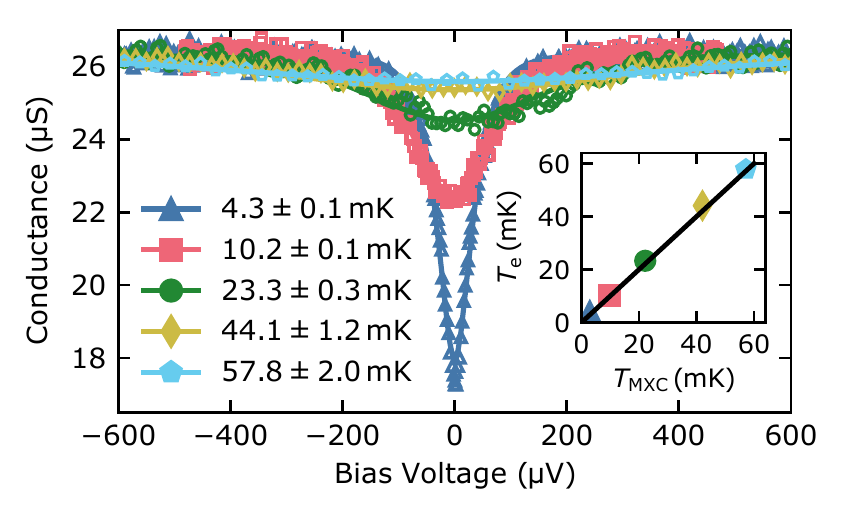}
    \caption{Calibration and measurement of the jCBT device. The CBT conductance dip is measured at five different temperatures. A full tunneling ME model is simultaneously fitted to the three highest temperature curves. The junction capacitance $C = \SI{185.7 +- 1.0}{\femto \farad}$ and the tunnel junction resistance $R_\mathrm{T} = 23.015 \pm 0.004 \, \mathrm{k \Omega}$ were common to all three fits while the four values of $T_e$ were allowed to differ. Following this, the coldest set of conductance data is fitted to the model using the previously determined parameters and allowing only the electron temperature to vary. This yielded a base electron temperature of \SI{4.3 +- 0.1}{\milli\kelvin}. The inset compares the CBT electron temperature $T_e$ with the dilution refrigerator mixing chamber temperature $T_\mathrm{MXC}$ (the solid line corresponds to $T_\mathrm{e}=T_\mathrm{MXC}$). There is a small amount of overheating of the CBT above the dilution refrigerator base temperature \SI{2.9 \pm 0.2}{\milli\kelvin}, as measured using a vibrating wire viscometer in the $\mathrm{^3He}$ -- $\mathrm{^4He}$ mixture~\cite{Bradley1990,Pentti2011}.}
    \label{fig:supp_jCBTcalibration}
\end{figure}

\begin{figure*}[htb!]
    \includegraphics{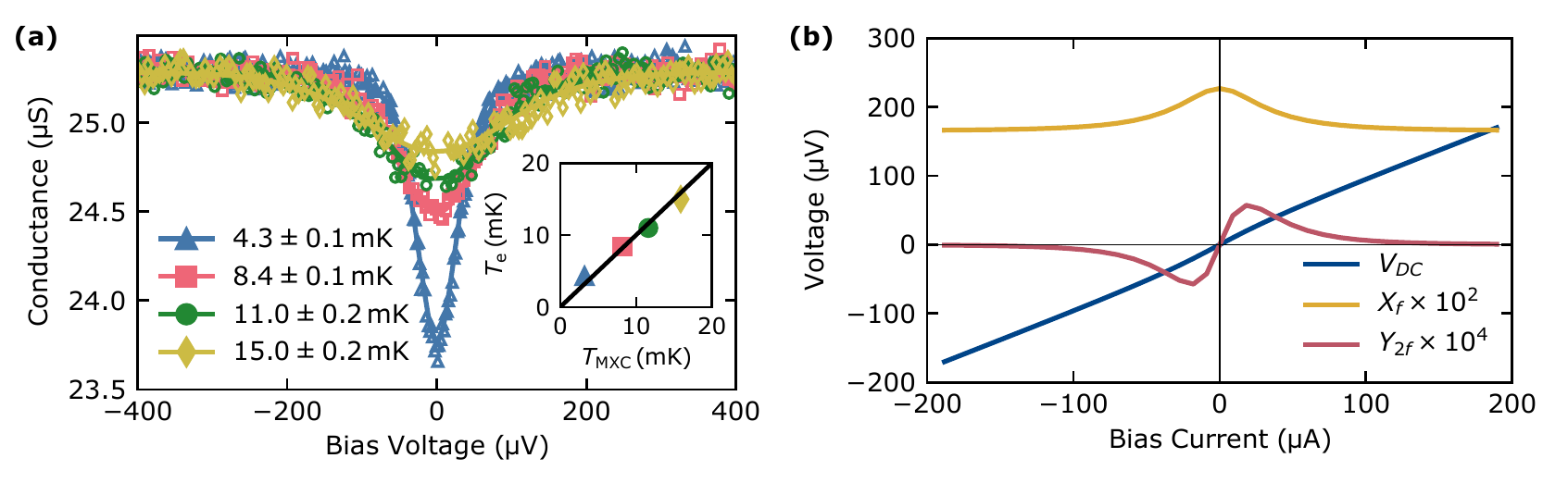}
    \caption{Calibration and measurement of the gCBT device. Panel (a) shows the CBT conductance dip measured at four different temperatures. Self-calibration using a jCBT model gives the parameters $C = \SI{1.155 +- 0.008}{\pico \farad}$ and $R_\mathrm{T} = \SI{23.9698 \pm 0.0008}{\kilo\ohm}$. While the ME model correctly predicts the conductance of a gCBT in the universal regime, it does not accurately capture the electrostatics of a gCBT. Therefore, the fitted value of $C$  reflects the charging energy and should be thought of as a scaling parameter rather than a specific physical capacitance in the device. Using the fitted parameters, we find that the coldest curve corresponds to an electron temperature of \SI{4.3 +- 0.1}{\milli\kelvin}. The inset in panel (a) compares the CBT electron temperature $T_e$ with the dilution refrigerator mixing chamber temperature $T_\mathrm{MXC}$. There is some overheating of the CBT above the dilution refrigerator base temperature $\SI{3.2 \pm 0.1}{\milli\kelvin}$, as measured using a vibrating wire viscometer operating in the dilute phase of the  $\mathrm{^3He}$ -- $\mathrm{^4He}$ mixture~\cite{Bradley1990,Pentti2011}. The solid line corresponds to $T_\mathrm{e}=T_\mathrm{MXC}$. Panel (b) illustrates how the center of the conductance dip is tracked in the secondary mode, i.e. when only measuring at zero bias. The dark blue line shows the CBT current-voltage characteristic. The slightly non-linear conductance of the CBT around zero-bias results in the curve $X_f$ seen at the AC excitation frequency $f$ and the curve $Y_{2f}$ at the second harmonic $2f$. The value of $Y_{2f}$ is positive for biases above zero and negative for biases below and so, when it is large enough to be measured, this signal is used in feedback to zero the DC current through the CBT.
    \label{fig:supp_gCBTcalibration}}
\end{figure*}

Figure~\ref{fig:supp_jCBTcalibration} shows the self-calibration of the jCBT device that was used to obtain the data shown in Fig.~\ref{fig:demag} of the article. Figure~\ref{fig:supp_gCBTcalibration}(a) shows the self-calibration of the gCBT device that was used to obtain the data shown in Fig.~\ref{fig:demag2} of the article. After calibration, the temperature of each CBT is determined by measuring its conductance at zero bias. The temperature is measured continuously during cooling experiments as the CBT is magnetised, pre-cooled, and demagnetised.

An experimental complication arises with very low temperature CBT measurements due to the narrowing of the conductance dip. For example, at \SI{1}{\milli\kelvin} our 33 junction arrays produce a conductance curve FWHM of \SI{15}{\micro\volt}. This means that a small DC offset of even $\sim\SI{100}{\nano\volt}$ can cause a significant pessimistic error in the measured electron temperature. To minimize this error, we track the center of the dip during the cooling process and correct any DC offset using feedback, as described in Fig.\@~\ref{fig:supp_gCBTcalibration}(b).

The intrinsic uncertainty in the temperature measured by a CBT is affected by nonuniformities in the tunnel junction resistances, tunnel junction capacitances and total island capacitances~\cite{Hirvi1995,Farhangfar1997,Hahtela2016,Pekola2022}. Moreover, at very low temperatures the conductance is affected by the unknown, static offset charge of each island~\cite{Hirvi1996,Yurttaguel2021}. For the devices studied here, the first contribution should be insignificant for the purpose of this work. The fabrication process is expected to produce very uniform tunnel junctions~\cite{Prunnila2010} and the CBT uncertainty is relatively insensitive to such variations. Furthermore, for all the data shown, the CBTs are operating at sufficiently high temperatures for the effect of offset charges to be neglected. This assumption is weakest for the coldest electron temperatures shown in Fig.~\ref{fig:demag}. However, using numerical results from Fig. 8 in~\cite{Yurttaguel2019}, we have confirmed that the scale of the uncertainty due to unknown offset charges at these temperatures is small. In practice, our temperature uncertainty is dominated by two things: at higher temperatures (in the universal regime) the uncertainty of fitting parameters in the calibration for operating in secondary mode is the greatest source of uncertainty. This is the origin of the temperature uncertainties shown in Fig.~\ref{fig:supp_jCBTcalibration} and Fig.~\ref{fig:supp_gCBTcalibration}. The same is true for uncertainties given in the main text. For the jCBT, at temperatures approaching 1\,mK and below, the electron temperature measurement is also affected by the amplitude of the AC excitation used to read out the differential conductance. 

\begin{figure}[htb!]
    \includegraphics[width=1\linewidth]{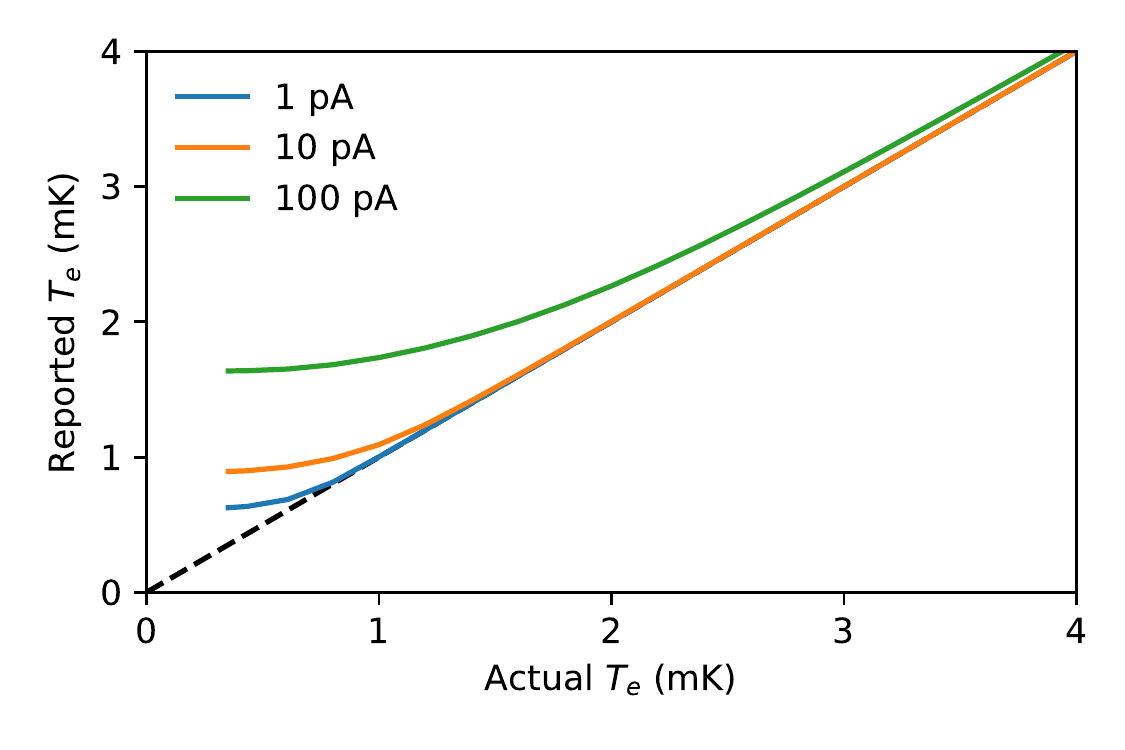}
    \caption{Junction CBT temperature readout saturation in the master equation model. For a decreasing actual electron temperature, the junction CBT temperature readout saturates at a level that depends on the AC excitation used to read out the CBT conductance. The excitation used in the experiment in Fig.~\ref{fig:demag} is the order of 10\,pA. For 10\,pA excitation and at $T_\mathrm{e}=0.35$\,mK, the "reported" electron temperature calculated from the measured jCBT conductance would be $\approx $ 0.9\,mK. The dashed line shows where the readout value is equal to the actual temperature.
    \label{fig:supp_jCBTsaturation}}
\end{figure}

We study the uncertainty due to finite AC excitation by sampling the theoretical CBT conductance peak in the master equation model with a simulated current excitation signal. The resulting voltage signal is analyzed to find the signal at the frequency of the excitation, simulating the operation of a lock-in amplifier. The resulting conductance reading, converted to "reported" electron temperature is shown in Fig.~\ref{fig:supp_jCBTsaturation}. For a 10\,pA excitation, as used in the experiments, the apparent electron temperature saturates around $T_\mathrm{e}\approx 0.9$\,mK. This can explain the readout saturation observed in the jCBT experiment in Fig.~\ref{fig:demag}. Reducing the AC current amplitude further is not possible in practice because the noise floor of the lock-in measurement would dictate an integration time that is too long for the timescale of the experiment. 

The agreement between the measured and predicted electron temperatures in Fig.~\ref{fig:demag} and the results in Fig.~\ref{fig:supp_jCBTsaturation} is surprisingly good, and one might be compelled to infer the actual electron temperature from conductance measurements at these temperatures by correcting for the saturation using Fig.~\ref{fig:supp_jCBTsaturation}. However, the master equation model is known to become increasingly unreliable at temperatures where the conductance suppression is stronger than $G_0/G_T \approx 0.65$~\cite{Feshchenko2013,Yurttaguel2021}. For this reason, the results in Fig.~\ref{fig:supp_jCBTsaturation} cannot be used to infer the actual electron temperature in Fig.~\ref{fig:demag}. Instead, the values reported in Fig.~\ref{fig:demag} represent a reliable upper limit for the electron temperature, with a lowest possible reading close to $1\,\mathrm{mK}$ due to the AC excitation.

\subsection*{Simulations}

The demagnetization performance of a copper island as shown in Fig.~\ref{fig:schematic}b of the main text is determined by the subsystem heat capacities and the thermal links between the subsystems. For the copper subsystem heat capacities we use theoretical temperature dependencies with material parameter values taken from the literature. The electron-nuclear and phonon-phonon couplings are described by first-principles expressions. The phonon-electron coupling is described by a phenomenological formula confirmed by independent prior experiments. The expressions and parameter values are explained in more detail in Ref.~\cite{jones2020progress}. 

The thermal subsystem couplings are shown in terms of thermal resistances $R$ in Fig.~\ref{fig:schematic}b. For numerical purposes it is however more convenient to express the couplings in terms of the heat flow $\dot{Q}$ between each subsystem pair. 

The most important of the subsystem couplings is that controlling the heat flow between the phonons and conduction electrons~\cite{Wellstood1994},

\begin{equation}
    \dot{Q}_\mathrm{pe} = \Sigma V (T_\mathrm{p}^5 - T_\mathrm{e}^5),
\end{equation}
where $\Sigma$ is a material specific constant ($\SI{2}{\giga\watt\per(\metre ^3\kelvin ^5)}$ for Cu~\cite{Viisanen2018,Meschke2004}), and $V$ the volume of the copper island. This rapid temperature dependence allows both efficient pre-cooling of the electrons by the phonons down to a few mK and almost total thermal decoupling of the electrons from the phonons when the electrons are cooled down by the nuclei during the demagnetization. 

The heat flow from the electrons to the nuclei reads 

\begin{equation}
    \dot{Q}_\mathrm{en} =\frac{\lambda_n n}{\mu_0 \kappa} \frac{B^2 + B_0^2}{T_\mathrm{n}} (T_\mathrm{e} - T_\mathrm{n}) = C_\mathrm{n} (T_\mathrm{e} - T_\mathrm{n}) T_\mathrm{n}/\kappa.
\end{equation}
Here $\lambda_n $ is the molar nuclear Curie constant of copper, $n$ is the number of copper moles, $\mu_0$ is the permeability of free space, $\kappa=\tau_1 T_\mathrm{e}$ is the Korringa constant for copper with $\tau_1=$\,\SI{1.2}{\kelvin \second}, and $B_0=$\,\SI{0.36}{\milli\tesla} is the residual dipole field in copper. The copper nuclear heat capacity $C_\mathrm{n}$ is defined below. 

Finally, the phonon-phonon heat flow between the silver platform and the copper phonons is determined by the boundary Kapitza resistance \cite{RevModPhys.61.605}

\begin{eqnarray}
    \dot{Q}_\mathrm{pp} &=& (T_\mathrm{p, Ag} - T_\mathrm{p, Cu})/R_\mathrm{pp}\\ 
    &=&    (T_\mathrm{p, Ag} + T_\mathrm{p, Cu})^2 (T_\mathrm{p, Ag}^2 - T_\mathrm{p, Cu}^2) A /(8 r_\mathrm{K}),
\end{eqnarray}
where $A$ is the contact area separating the two phonon systems, $r_\mathrm{pp} \equiv R_\mathrm{pp}  A T^{3}  \sim10^{-2}$\SI{}{\kelvin^4 \meter^2/\watt} for a multi-layer metal-insulator interface as detailed in the next section, and we used the approximation that $T=(T_\mathrm{p, Cu}+T_\mathrm{p, Ag})/2$. 

\begin{figure}
\includegraphics[width=1\linewidth]{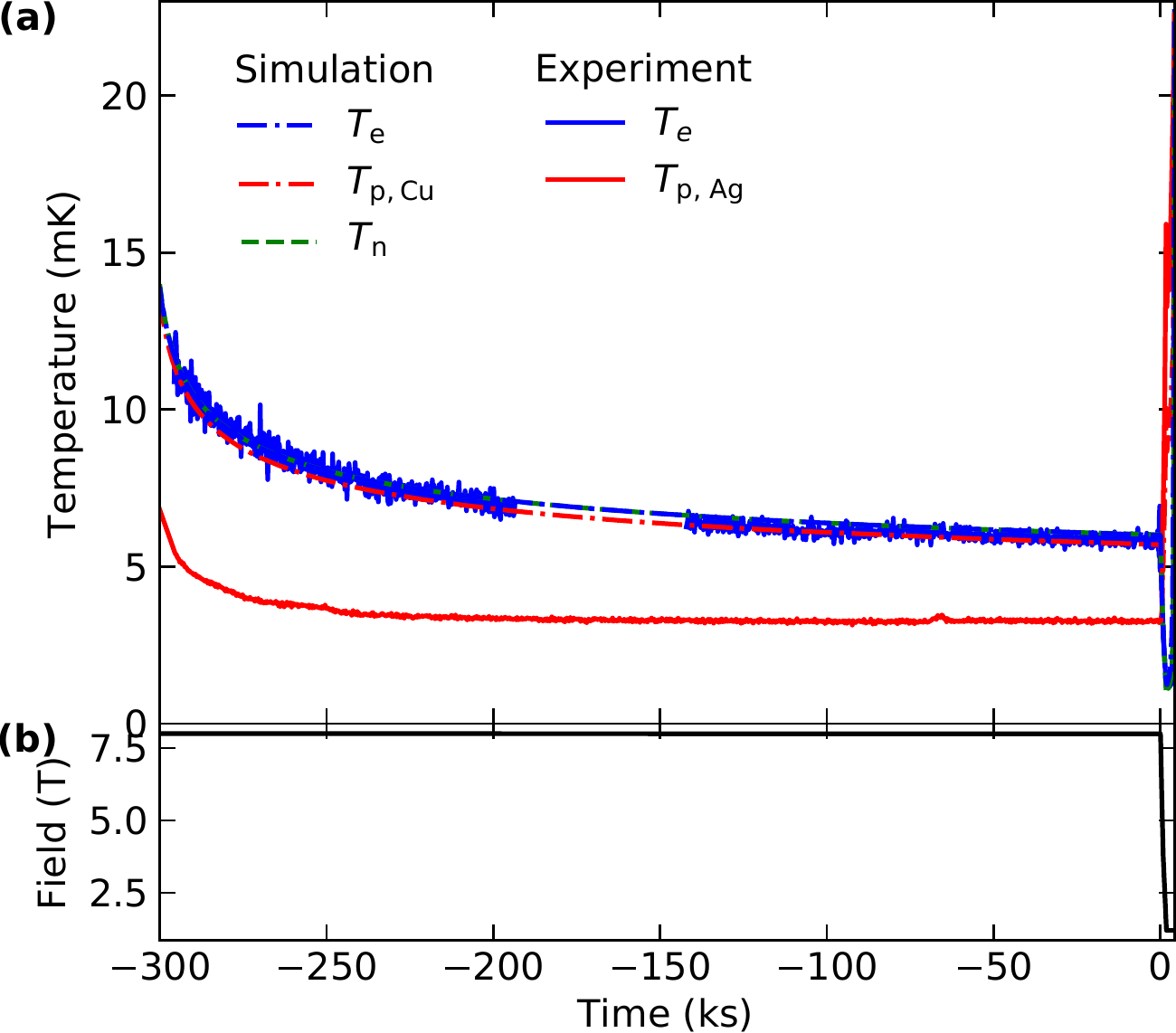}%
\caption{Pre-cooling of the gCBT. (a) The measured measured mixing chamber temperature (red solid line) decreases rapidly in the beginning of the pre-cooling and then levels off at 3\,mK. The measured electron temperature is shown by the blue solid line, segments of which are removed to highlight the simulated temperature evolution. The simulated temperature differences between the copper subsystems are negligible. Thus, the only thermal coupling that the pre-cooling rate is sensitive to is $R_\mathrm{pp}$, which can be fitted to the early part of the data at $t<-200$\,ks where $Q_0$ is negligible, yielding $R_\mathrm{pp}=0.8 \times 10^{-2}$\SI{}{\kelvin^4 \meter^2/\watt}. The temperature at which the pre-cooling levels off is determined the constant heat leak $Q_0$. We find $Q_0\approx\,$\SI{0.1}{\femto\watt} by fitting the late part of the data. (b) The magnetic field is set to \SI{8}{\tesla} during the refrigerator pre-cooling at $t<0$.\label{fig:demag2_precool}}
\end{figure}

To simulate the subsystem temperature evolution, we also need the heat capacities of the subsystems \cite{jones2020progress}. The copper electrons have the heat capacity

\begin{equation}
    C_\mathrm{e} = n c_\mathrm{e} T_\mathrm{e},
\end{equation}
where $c_\mathrm{e}=0.691\times10^{-3}$\SI{}{\joule /( \kelvin^2 \mol)}. The copper nuclear heat capacity reads

\begin{equation}
        C_\mathrm{n} = n c_\mathrm{n} (B^2 + B_0^2) /T_\mathrm{n}^2,
\end{equation}
where $c_\mathrm{n}=3.22 \times 10^{-6}$\SI{}{\joule \kelvin /(\tesla^2 \mol)}. Finally, copper phonons have the heat capacity

\begin{equation}
    C_\mathrm{p} = n c_\mathrm{p} T_\mathrm{p}^3,
\end{equation}
where $c_\mathrm{p}=4.73\times10^{-5}$\SI{}{\joule /(\kelvin^3 \mol)}.

In a static magnetic field the copper subsystem temperatures are  governed by a set of three differential equations:

\begin{eqnarray}
\frac{\mathrm{d} T_\mathrm{p}}{\mathrm{d} t} &= (\dot{Q}_\mathrm{pp} - \dot{Q}_\mathrm{pe})/C_\mathrm{p},& \\
\frac{\mathrm{d} T_\mathrm{e}}{\mathrm{d} t} &= (\dot{Q}_\mathrm{pe}-\dot{Q}_\mathrm{en} + Q_0)/C_\mathrm{e},& \\
\frac{\mathrm{d} T_\mathrm{n}}{\mathrm{d} t} &= \dot{Q}_\mathrm{en}/C_\mathrm{n}.&
\end{eqnarray}
The entropy of the nuclear system is purely a function of $B/T_\mathrm{n}$, and for an ideal adiabatic demagnetization process this ratio will remain constant~\cite{Pobell2007}. That is, $T_\mathrm{n}$ decreases as proportional to the change in the field magnitude assuming the ratio of $\dot{Q}_\mathrm{en}$ and $ C_\mathrm{n}$ gives small changes as compared with the demagnetisation rate. This condition is satisfied in all the simulations presented in this Article unless mentioned otherwise.

We solve the temperature evolution in the copper subsystems in three steps using the Python partial differential equation solver package Scripy.integrate.odeint. First, the stable pre-cooling state is fitted using $Q_0$ as a fitting parameter. The demagnetization ramp is simulated by reducing $B$ by a small step and adjusting $T_\mathrm{n}$ and $C_\mathrm{n}$ correspondingly and instantaneously, followed by solving the temperature evolution of the copper subsystems for the time that corresponds to this tiny step in $B$. This approximation is justified provided the nuclear temperature is not significantly affected by the heat leak from the electrons during the demagnetization. That is, this approximation may overestimate the cooldown speed and minimum nuclear temperature reached if the nuclear heat capacity is being exhausted while the magnetic field is still changing. However, this is not the case in any of the simulations presented in this paper. The stepping procedure is repeated until the final magnetic field is reached. Third, the constant-field evolution of the system is solved until the total hold time is revealed. 

\subsection{Kapitza resistance from a layered substrate}

The substrate that the copper demagnetization islands are attached to has the following layered structure. The \SI{6.5}{\micro \meter}-thick layer of copper is followed by a layer of aluminum (\SI{0.25}{\micro \meter}), a layer of silicon oxide (\SI{0.25}{\micro \meter}), silicon (\SI{330}{\micro \meter}) and silver. The silver layer is well thermalized to the mixing chamber of the refrigerator. Thus, phonons in the silver layer are at temperature $T_\mathrm{MXC}$. 

The total resistance for phonon heat flow, $R_\mathrm{pp}$, comes from the combined effect of all the interfaces between the copper and the silver layers. Due to the tiny volume of the aluminum layer, essentially all electron-phonon heat flow in this system takes place in the copper volume. The phonon heat flow thus needs to pass through four interfaces between silver and copper, each of which has its own Kapitza resistance. For the silver-silicon interface $R_\mathrm{K}  A T^{3} \approx 1.4 \times 10^{-3}$\SI{}{\kelvin^4 \meter^2/\watt} \cite{RevModPhys.61.605}. Assuming the other interface types present, not discussed in Ref.~\cite{RevModPhys.61.605}, carry Kapitza resistances of similar order of magnitude as the interface types listed, the total combined thermal resistance becomes $R_\mathrm{pp}  A T^{3} \sim 10^{-2}$\SI{}{\kelvin^4 \meter^2/\watt}. Here we assumed the thermal resistances can be summed up, which is justified as long as the heat capacities of the substrate layers are negligible and the substrate is mostly covered by the copper islands (the substrate contact area with the silver layer is roughly equal to the area covered by the islands). Note that the intrinsic thermal resistance of even the thickest intermediate layer, silicon, is orders of magnitude smaller than the interface resistances and can thus be neglected (see Fig.~5 in \cite{klitsner1987phonon}).

We can extract the combined boundary resistance for each CBT device by fitting pre-cooling data using the effective Kapitza resistance as a fitting parameter. Fig.~\ref{fig:demag2_precool} shows four days of measured temperatures including the demagnetization shown in detail in Fig.~\ref{fig:demag2}. In the simulation the only significant temperature step is that between $T_\mathrm{p,Cu}$ and $T_\mathrm{p,sbstr}$. Therefore, the initial slope of the temperature decrease in copper is controlled by $R_\mathrm{pp}$ only. The fitted value for the gCBT is $R_\mathrm{pp}=0.8 \times 10^{-2}$\SI{}{\kelvin^4 \meter^2/\watt} and for the jCBT $R_\mathrm{pp}=1.5 \times 10^{-2}$\SI{}{\kelvin^4 \meter^2/\watt}. At the end of the pre-cooling process the temperature saturates at a value which depends on the constant heat leak $Q_0$. The fitted value is $Q_0=0.1$\,fW for both devices.

\begin{figure}
\includegraphics[width=1\linewidth]{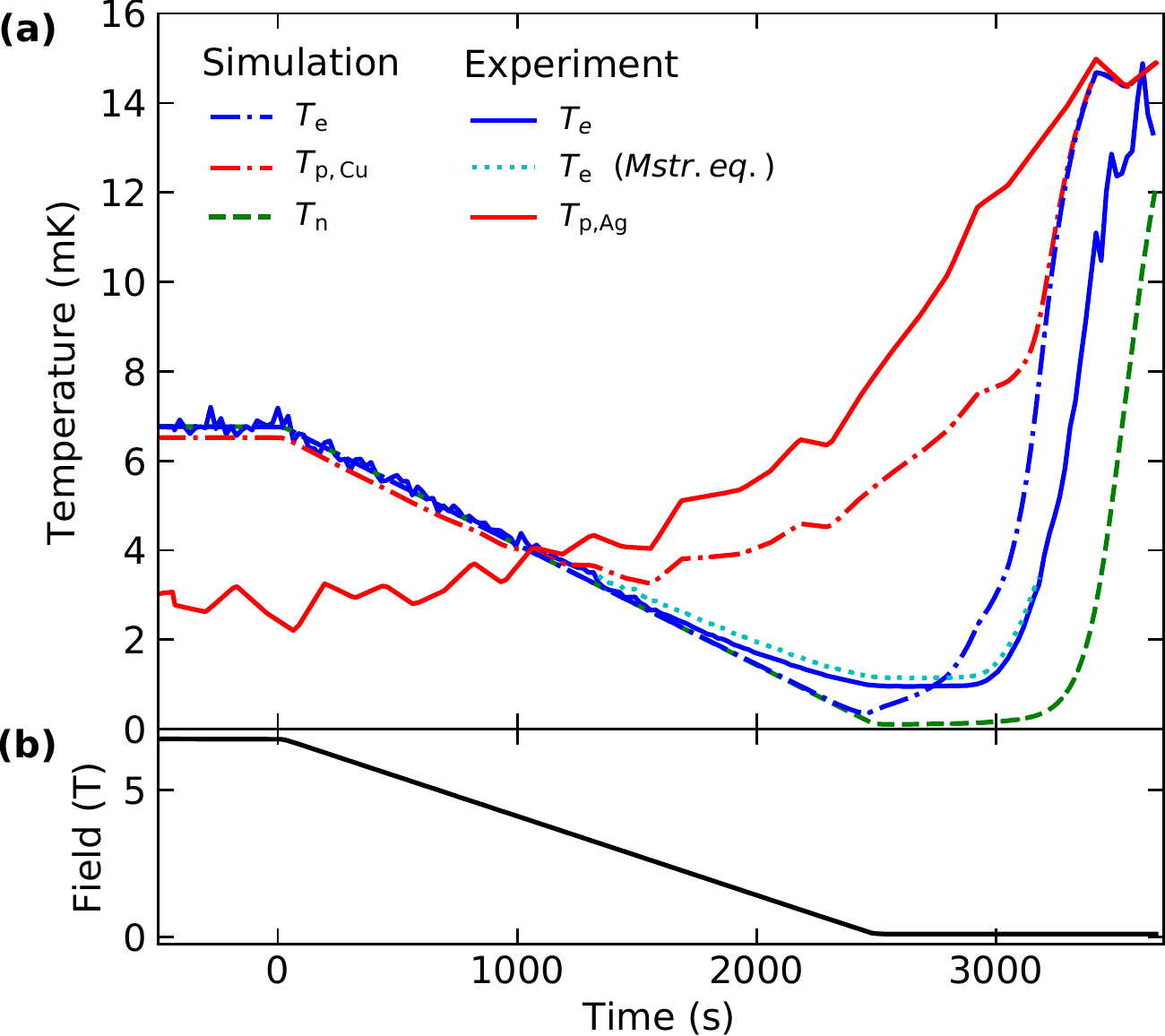}%
\caption{Junction CBT demagnetization experiment and simulation with perturbed Kapitza resistance. (a) The measured jCBT electron temperature is shown as explained in \ref{fig:demag} in the main Article. In the simulation, we used a perturbed value for the  total Kapitza resistance, $R_\mathrm{pp}=1.0 \times 10^{-2}$\SI{}{\kelvin^4 \meter^2/\watt}. The resulting hold time is about 100\,s shorter than with a 50\% higher Kapitza resistance. Otherwise the temperature evolution is very similar with that shown in Fig.~\ref{fig:demag}. The fitted constant heat leak is $Q_0\approx$\,\SI{0.1}{\femto\watt}. (b) The magnetic field is set to \SI{6.75}{\tesla} during the pre-cooling at $t<0$, and is then reduced linearly to \SI{100}{\milli\tesla} in 2500\,s. \label{fig:demag_speculation_varyK}}
\end{figure}

The difference in the fitted values of $R_\mathrm{pp}$ is explained by the different device geometries: The gCBT is otherwise identical with the jCBT, but it is covered with a layer of dielectric material (\SI{0.4}{\micro\meter}) followed by a layer of metal \cite{jones2020progress}. The top metal layer is also well thermalized to the mixing chamber. This doubles the effective contact area between the gCBT phonons and the nearest metal at $T_\mathrm{MXC}$, which matches with the halved $R_\mathrm{pp}$ measured for this device. We note that this result also means that adding a dielectric layer over the device can be used to tune the Kapitza resistance and thus also the pre-cooling performance. The geometry difference does not affect the heat leak, consistent with the fitted values above. 

We emphasize that the demagnetization performance of either device only weakly depends on the precise value of $R_\mathrm{pp}$. This is because both the electron-phonon coupling within the copper and the phonon-phonon coupling via $R_\mathrm{pp}$ depend strongly on the interface temperature differences. That is, a tiny temperature change in a given steady-state configuration corresponds to a large change in the coupling magnitude. To demonstrate that the demagnetization outcome is not sensitive to the precise value of $R_\mathrm{pp}$ beyond the rough estimate that $R_\mathrm{pp}  A T^{3} \sim 10^{-2}$\SI{}{\kelvin^4 \meter^2/\watt}, Fig.~\ref{fig:demag_speculation_varyK} shows a simulated jCBT demagnetization with $R_\mathrm{pp}=10^{-2}$\SI{}{\kelvin^4 \meter^2/\watt} instead of $R_\mathrm{pp}=1.5 \times 10^{-2}$\SI{}{\kelvin^4 \meter^2/\watt}. As compared with the simulation shown in Fig.~\ref{fig:demag}, the hold time decreases by about 180\,s. This is consistent with the decreased thermal isolation from $T_\mathrm{MXC}$. Otherwise the simulation outcomes are nearly identical.

\subsection{Engineered Kapitza resistance for 10\,mK operation}

\begin{figure}[ht!]
\includegraphics[width=1\linewidth]{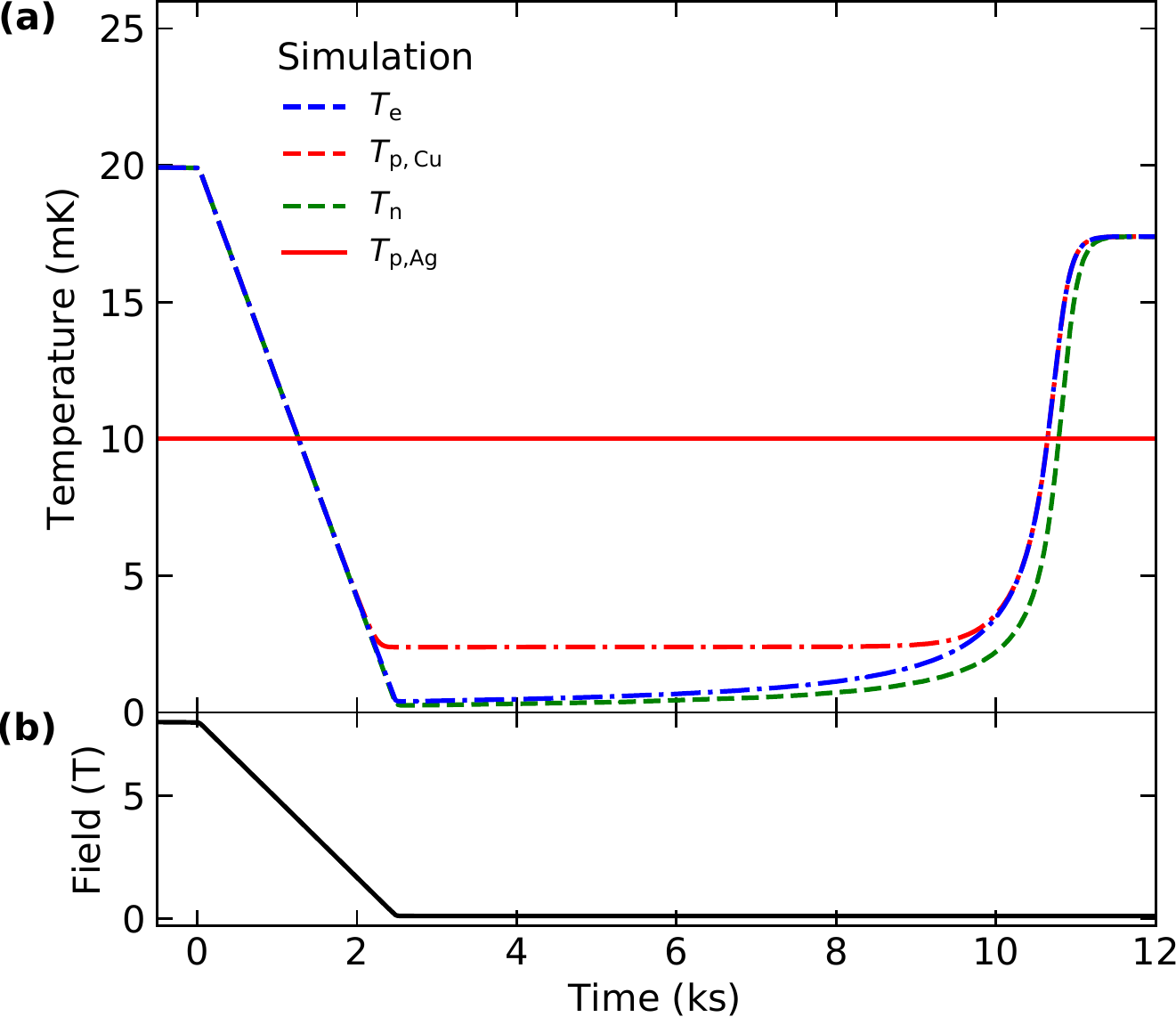}%
\caption{Junction CBT demagnetization simulation with Kapitza resistance increased by two orders of magnitude. (a) The measured jCBT electron temperature is shown as explained in \ref{fig:demag} in the main Article. In the simulation, we used a perturbed value for the  total Kapitza resistance, $R_\mathrm{pp}=1.0$\,\SI{}{\kelvin^4 \meter^2/\watt}. The resulting minimum electron temperature is 0.4\,mK and the hold time below one mK is well over an hour. The constant heat leak used is $Q_0\approx$\,\SI{0.1}{\femto\watt}. (b) The magnetic field is set to \SI{8}{\tesla} during the pre-cooling at $t<0$, and is then reduced linearly to \SI{100}{\milli\tesla} in 2500\,s. \label{fig:demag_speculation_highK}}
\end{figure}

It is possible to engineer the Kapitza resistance to improve device performance. In Fig.~\ref{fig:demag_speculation_highK} we have increased the simulated Kapitza resistance by two orders of magnitude. This allows reaching and holding sub-mK temperatures even with the silver phonon temperature as high as 10\,mK and the pre-cooled CBT temperature 20\,mK. These are typical performance figures in a commercial cryogen-free refrigerator.

Such a large total Kapitza resistance could be created by increasing the number of thin layers that make the substrate to several hundred or by decreasing the contact area between the substrate and the silver layer underneath using a suitable spacer between them. A sophisticated alternative would be to use a suitable superconducting metal layer in between the silver and the substrate with a superconducting transition temperature below the full field during precooling but above the final field. This layer would act as a heat switch, disconnecting the device thermally from the refrigerator during the demagnetization. As an example, TiN has a superconducting critical field of 5\,T \cite{pierson1996handbook}.

\end{document}